\documentstyle[11pt]{article}
\font\mtbb=msbm10 scaled \magstep1
\def\R{\hbox{\mtbb R}}

\def\Z{\hbox{\mtbb Z}}
\def\CA{{\cal A}}
\def\CM{{\cal M}}
\def\CF{{\cal F}}

\def\CR{{\cal R}}
\def\r{\rho}
\def\s{\sigma}
\def\k{\kappa}
\def\r{\rho}
\def\m{\mu}
\def\n{\nu}
\def\a{\alpha}
\def\b{\beta}
\def\g{\gamma}

\def\implies{\Rightarrow}
\def\bt{\bullet}
\def\pt{\partial}

\def\GH#1#2#3{\Gamma^{#1}_{#2#3}}
\def\B2#1#2{B_{#1#2}}
\def\OD{\O{1} \tsa \O{1} }

\def\GT#1#2{\Gamma^{#1}_{#2}}
\def\t#1#2{\theta_{#2}^{#1}}

\catcode`@=11
\def\ra#1{\;\; \vbox{\m@th\ialign{##\crcr
     \hfil ${}^{\scriptstyle #1}$\hfil\crcr\noalign{\kern-\p@\nointerlineskip}
      $\hfil\rightarrow\hfil$\crcr}} \;\;}

\def\ri#1{\;\; \vbox{\m@th\ialign{##\crcr
     \hfil ${}^{\scriptstyle #1}$\hfil\crcr\noalign{\kern-\p@\nointerlineskip}
      $\hfil\hookrightarrow\hfil$\crcr}} \;\;}

\catcode`@=12 

\def\<{\langle}
\def\>{\rangle}
\def\nn{\nonumber}

\def\ts{\mathop{\otimes}}
\def\tsa{\otimes_{\scriptstyle \cal A}}
\def\id{\hbox{id}}

\def\beq{\begin{equation}}
\def\eeq{\end{equation}}
\def\bea{\begin{eqnarray}}
\def\eea{\end{eqnarray}}

\def\l#1{\label{#1}}

\def\={\; = \;}

\def\om{\omega}
\def\id{\hbox{id}}
\def\kr{\hbox{Ker\ }}

\def\O#1{\Omega^{#1}}
\def\pt{\partial}
\def\bt{\wedge}
\def\={\; = \;}

\def\bo{\begin{observation}}
\def\eo{\end{observation}}

\begin{document}
\thispagestyle{empty}

\null
\vspace{5cm}

\def\thefootnote{\fnsymbol{footnote}}
\begin{center}
{\LARGE On Some Aspects of Linear Connections \\
in Noncommutative Geometry} \\

\ \\
\ \\

Andrzej Sitarz \footnote{Partially supported
by KBN grant 2P 302 103 06 } \\
\ \\
{\em Department of Theoretical Physics \\
Institute of Physics, Jagiellonian University \\
Reymonta 4, 30-059 Krak\'ow, Poland \\
e-mail: sitarz@if.uj.edu.pl}
 \\

\end{center}

\vfill
\begin{abstract}

We discuss two concepts of metric and linear connections
in noncommutative geometry, applying them to the case
of the product of continuous and discrete (two-point) geometry.

\end{abstract}

\vfill
\noindent{\sc TPJU 5/95}\\
{\sc February 1995 }

\vfill
\newpage

\section{Introduction}

Noncommutative geometry  [1-5] is one of the most attractive mathematical
concepts in physics that could be applied in fundamental field theory.
So far, the investigations of gravity in this framework have been
concentrated on the case of the product of Minkowski space by a two-point
space, which has been motivated by the Standard Model (see  [6-8])

Their methods,
however, did not use the whole structure of noncommutative geometry,
in particular, the definitions of metric and linear connections did not
use the bimodule structure of differential forms.

Only recently some general ideas concerning linear connection and
metric have been proposed and discussed for other examples . They [9-11]
are based on the idea that a key role in the introduction of these
structure plays as generalised permutation operation.

A different model of the generalisation of the metric as well as a simple
model of gravity on the product of Minkowski space and two-point space has
been already discussed by us earlier, with some encouraging results \cite{JA}.

In this paper, we shall discuss two methods of construction of the metric
and linear connections based on two different concepts, first as proposed
in  [9-11], based on symme tric metric and bimodule property of
linear  connection,
the other one, which uses hermitian metric and left-linearity of linear
connection and follows the idea of our previous paper (though it differs
in few significant points). We shall try to derive the
consequences of these models
for the considered example. Our main aim is to determine what
conditions are necessary, what could be abandoned and what are too
strict for noncommutative geometry. Of course, the basic test is the
agreement with the standard differential geometry.

\section{Notation}

Our basic data is a (graded) differential algebra $\O{}$ with the
external derivative $d$ obeying the graded Leibniz rule:
\beq
d( u \bt v) = du \bt v  + (-1)^{\hbox{\footnotesize deg\ }u} u \bt dv,
\eeq

We shall denote by $\O{n}$ a bimodule of $n$-forms, $n \geq 1$ and
we shall write ${\CA}$ for $\O{0}$.

Let $\pi_n$ be the canonical projection $\pi_n: \O{\ts n} \to \O{n}$,
$n \geq 2$, for simplicity we shall often write $\pi$ unless it is
necessary to specify the index $n$.
We assume also that our external algebra is a graded $\star$-algebra
and we have
\beq
d (\omega^\star) \= (d \omega)^\star,
\eeq

To end this section let us remind the basic notation of an example
of a noncommutative differential calculus on a product of $\R^n$ and
a two-point space.

The algebra ${\CA}$ contains of functions on $R^n\times \Z_2$,
with pointwise addition and multiplication (also with respect to the
discrete coordinates). The bimodule of one-forms is generated by $n+1$
elements: $\{ dx^i\}_{i=1,\ldots,n}$ and $\chi$, with the following set
of multiplication properties:

\bea
f(x,p) dx^i & = & dx^i f(x,p) \\
f(x,p) \chi & = & \chi f(x,-p)
\eea
where $p$ denotes the discrete coordinate taking values $+$ and $-$.

The external derivative is defined as follows:

\beq
d f(x,p)  \= \sum_{i=1}^n dx^i \pt_i f(x,p) + \chi \pt f(x,p)
\eeq

where $\pt_i$ is the usual partial derivative and
$\pt f = (1 - \CR)f$, $\CR$ being the morphism, which
flips the discrete coordinate: $\CR f(x,p) =  f(x, -p)$.

The external algebra is built with the following multiplication rules:

\bea
dx^i \bt dx^j & = & - dx^j \bt dx^i, \\
dx^i \bt \chi & = & - \chi \bt dx^i, \\
d( \chi) & = & 2 \chi \bt \chi,
\eea

and is infinite-dimensional, as $\chi \bt \chi$ does not vanish. One can
introduce a $\star$-algebra structure on this algebra, assuming that:
\bea
(dx^i)^\star \= dx^i, \;\;\;\;\;\; \chi^\star = - \chi. \l{conj1}
\eea

The differential calculus constructed in the above described way is just
a tensor product of external algebras on the continuous space (which is
a standard one) and the discrete two-point space (which is an universal
differential calculus).

\section{Symmetrization and antisymmetrization}

In the classical differential geometry the external algebra is defined
as an antisymmetrization of the tensor algebra of one-forms, therefore
these operations precede the construction of differential calculus.
In noncommutative geometry this situation could be different and we may
choose between several possibilities, all of them coinciding  in the case
of commutative differential structures.

\subsection{Antisymmetrization}

We may choose a similar way as in the standard differential geometry and,
having constructed the first order differential calculus,
(i.e. bimodule $\O{1}$ and $d : \CA \to \O{1}$, which obeys the Leibniz
rule) we may look for a bimodule isomorphism $\sigma$:
\beq
\s: \OD \to \OD,
\eeq
which would correspond to the permutation: $dx^a \tsa  dx^b \to dx^b \tsa
dx^a$.

Then, we define the noncommutative analogue of the symmetrizing morphism
on $\OD$ as $1 - \s$ and, consequently, the bimodule of two-forms as
a quotient bimodule $\OD / S$, where $S = \kr (1-\s)$. However,
we must ensure that the following consistency conditions hold: for any
 elements $a_i,b_i \in \CA$ we must have:
\beq
\sum_i a_i db_i = 0 \; \implies \sum_i d a_i \tsa d b_i \in  S.
\eeq

If $\sigma$ satisfies a braid group relation on $\Omega^{\tsa 3}$ then
the construction of the whole differential algebra follows directly,
let us stress that it is not necessary to require $\sigma^2=1$.

\subsection{Symmetrization}

Having defined $\s$ and the external algebra we might define a symmetrization
morphism on $\OD$ as $1+\s$, however, we cannot guarantee without some
additional assumptions that:
\beq
\pi \circ (1+\s) = 0, \l{sym}
\eeq
Indeed, since $\kr \pi = \kr (1-\s)$, if $(1+\s) \xi \in \kr \pi$ we would have
that
either $\s \xi = -\xi$ or $(1-\s)(1+\s) \xi = 0$, so that $\s^2 \xi = \xi$
in both cases. Therefore, $\s^2=1$ is a necessary requirement (it is obvious
that it is also sufficient) for (\ref{sym}) to hold.

Another option (which has been discussed by \cite{MAD1}) is to assume the
existence
of $\s$ and the (\ref{symm}) relation without deriving the external calculus
{}from $\s$, in that case, however,  we can lose strict relations between
the calculus and $\s$, and the choice of $\s$ could be rather ambiguous.

\subsection{Symmetrization and Antisymetrization - All In One}

In what follows we shall discuss a possibility of deriving the
symmetrization and antisymmetrization operations from the external
algebra itself. Of course, without some additional assumptions this
is not possible, however, as one could see that these assumptions
are rather natural, we shall present the idea here.

Let $S$ denote $\kr \pi$ and $j$ be the inclusion of $S$ in $\OD$.
Then the following is a short exact sequence of bimodules over $\CA$:
\beq
0 \ri{} S \ri{j} \OD \ra{\pi} \O{2} \ra{} 0.
\eeq

If $\O{2}$ is a projective module the above exact sequence is a split
sequence, i.e. there exist maps $r$ and $\rho$:
\bea
r: & \OD \to S &   r \circ j = \id_S \\
\rho: &  \O{2} \to \OD &   \pi \circ \rho = \id_{\OD}
\eea
and, moreover:
\beq
\OD \cong S \oplus \O{2}. \l{split}
\eeq

The latter allows us to introduce a natural symmetrization and
antisymmetrization operations on $\OD$. For every $\xi \in \OD$
we can represent it as a sum $\xi_s+ \xi_a$, where $\xi_s \in j(S)$
and $\xi_a \in \rho(\O{2})$. Then the following map:
\beq
\s: \;\;\; \s(\xi) = \xi_s - \xi_a,
\eeq
is a bimodule homomorphism such that $\kr \s = 0$ and $\s^2=1$. One
can easily verify that $1-\s$ is then a projection on $\rho(\O{2})$
and $1+\s$ is a projection on $j(S)$.


\subsection{Example}

In the example discussed in this paper the situation is rather simple,
as the only nontrivial noncommutative part comes from the discrete geometry.
As all bimodules $\O{n}$ are free, we may use results of the last section. We
shall write here only the resulting homomorphism $\s$:
\bea
\s (dx^i \tsa dx^j) & = & dx^j \tsa dx^i \\
\s (dx^i \tsa \chi) & = & \chi \tsa dx^i \\
\s (\chi \tsa \chi) & = & - \chi \tsa \chi
\eea

\section{Metric}

The construction of metric is one of the most important issues in
noncommutative geometry. First, it is required in the studies of
field theories (in particular gauge theories) in this framework,
secondly it is a crucial step towards the analysis of gravity. We
shall outline here the commonly used definition and discuss
several points, which are still not well established.

\subsection{Definition}

It has been almost generally agreed that the proper generalisation of
the metric tensor is a bimodule map:
\beq
g: \;\; \OD \; \to \CA, \l{defi}
\eeq

as it is a natural extension of the standard bilinear map to the
noncommutative situation. If our differential algebra is a $\star$-algebra
one should also postulate:

\beq
g(u^\star,v^\star) \=  g(v,u)^\star, \l{herm}
\eeq

which guarantees that $g(u,u^\star)$ is self-adjoined.

The above mentioned properties of  the metric tensor translate
easily from the standard differential geometry into the noncommutative
geometry, however, the problems start, when we begin to analyse other
features of the metric tensor.

\subsection{Symmetry}

In the standard differential geometry one postulates that the metric
is symmetric, i.e. $g(u,v)=g(v,u)$ for any one-forms $u,v$. Of course,
this requirement cannot hold in noncommutative geometry, however, one could
think of replacing it by a different one, which  recover this property
in the commutative limit.

The ambiguity comes from the fact that even in classical geometry
one may look at this property of the metric from two different points.
First, one may view the symmetry as related to the hermitian metric
condition, then the appropriate generalisation should be just
(\ref{herm}). Another point of view relates the symmetry of the metric
to the symmetrization operation on the bimodule $\OD$, then the
corresponding generalisation should take the form \cite{MAD1}:
\beq
 g \circ \s \= g, \l{symm}
\eeq
where $\s$ is the bimodule isomorphism discussed earlier. We shall
now investigate the consequences of each of these definitions in
our example.

\subsubsection{Symmetric metric - example}

{}From the definition (\ref{defi}) we immediately get that the metric
evaluated on the generating one forms must be:
\bea
g( dx^i \tsa dx^j) & = & g^{ij} \l{met1}\\
g( dx^i \tsa \chi) & = & 0 \\
g( \chi \tsa dx^i ) & = & 0 \\
g(\chi \tsa  chi) & = & g \l{met4}
\eea
so that the 'mixed' components must vanish and $g^{ij}$,$g$ denote
the nonzero elements of the algebra $\CA$.

The hermitian metric condition (\ref{herm}) together with (\ref{conj1})
relations gives us:
\bea
( g^{ij} )^\star & = & g^{ji}, \\
  g^\star & = & g.
\eea

If we require the additional symmetry (\ref{symm}), we obtain:
\bea
 g^{ij}  & = & g^{ji}, \\
 g  & = & - g. \l{disy}
\eea
so that $g^{ij}$ is a real and symmetric tensor and $g$ vanishes. The
latter property is rather inconvenient and we shall now generalise
it and discuss in details.

\newpage
\subsection{Metric on Universal Differential Calculus}

So far, we have encountered a problem with the existence of
a (nontrivial) metric on the discrete space of two-points, if
we assumed its symmetry (\ref{symm}). This feature appears every
time we have an universal differential calculus:

\bo{1} {\em If $\O{}$ is an universal differential calculus, then
there exists no nontrivial metric on $\OD$, symmetric in the sense of
(\ref{symm}).} \\

{\bf Proof:} If the calculus is universal, then $\pi_n=\id_{\O{n}}$
and therefore $\s = - \id$. From (\ref{symm}) follows that $g = -g$,
hence $g \equiv 0$.
\eo

Such consequence is rather an undesired one, as one of its aftermath
would be the elimination Higgs-field components of the Standard Model
Lagrangian, as we shall see later.
Therefore, we should rather stick to the basic interpretation of
the symmetry property (\ref{herm}) of the metric.

\subsection{Metric on higher order forms}

Another standard property of the metric is the possibility of
extending its definition for modules of higher-order forms. We
shall propose here a scheme for generalisation of it in noncommutative
geometry. First, we shall extend $g$ to the tensor products of $\O{1}$

\bo{2} {\em The metric $g$ extends to
$\Omega^{\tsa n} \tsa \Omega^{\tsa n}$ in the following way: }
\bea
g( u_1 \tsa \ldots \tsa u_n \tsa v_1 \tsa \ldots \tsa v_n) \= \\
 \;\;\;\;\; \= g\left(u_1 \left( \ldots g(u_{n-1} g(u_n \tsa v_1)\tsa
v_2\right)
 \tsa \ldots \tsa v_n \right),
\nn
\eea
which  satisfies the basic requirements (\ref{defi}-\ref{herm}).
\eo

Now, using the result (\ref{split}) we may extend the metric for
higher order forms using the embedding $\rho$. For instance, in the case of
two-forms this would be:
\beq
g(\omega,\eta) \= g\left( \rho(\omega), \rho(\eta) \right). \l{mehi}
\eeq
for any two-forms $\omega,\eta$.

\subsubsection{Example - metric on two-forms}

Here we shall demonstrate how the metric acts on an arbitrary
two-form of our product geometry $\CF$:
\beq
\CF \= dx^i \bt dx^j F_{ij} + dx^i \bt \chi \Psi_i + \chi \bt \chi \Phi,
\eeq

Using the form of the metric (\ref{met1}-\ref{met4}) and the definition
(\ref{mehi}) we find:
\bea
g(\CF^\star, \CF) \= & - & F_{ij}^\star F_{kl} g^{ik} g^{kl} \\
& + & \frac{1}{2} \Psi^\star_i \Psi_j g (\CR (g^{ij}) + g^{ij}) \\
& - & \Phi^\star \Phi g \CR(g).
\eea

The first term is a standard one, coming only from the part of
continuous geometry, whereas the last comes only from the discrete
geometry and the middle one is mixed. Had we assumed the symmetry
condition (\ref{symm}) to hold, we would have had consequently $g=0$
and both additional terms that have origins in
discrete geometry would not appear. This would have profound
consequences for physics, as any field theory, and gauge theory
in  particular, would not feel the presence of discrete geometry
(apart from the simple fact that we would have had two seperate
copies of each field). In such situation no Higgs-type model would
be possible to obtain from the noncommutative geometry based on the
product of $\R^n \times \Z_2$, and one should look for models, which
involve products of differential calculi, which are not universal, to
obtain nontrivial results.

\section{Linear Connections}

As the standard methods of differential geometry use rather the
language of vector fields than differential forms, the translation
of the concept of linear connection is a delicate problem. We might
also look at the formulation of gauge theory in noncommutative geometry
to guess the best definition, let us remind that for any left-module
$\CM$ over $\CA$ the covariant derivative $D$ is defined as a map
$\CM \to \O{1} \tsa \CM$ such that:

\beq
D(a m) = da \tsa m + a D(m), \l{codi}
\eeq

which could be then extended to the degree 1 operator
$D: \Omega \tsa \CM \to \Omega \tsa \CM$.

One could easily apply this definition for the case of linear
connections (and the related covariant derivative) by replacing
$\CM$ with the appropriate object in this case, the bimodule
$\O{1}$. The problem starts when we begin to look at the bimodule
structure of $\O{1}$ and ask how $D$ acts on $u a$, $u \in \O{1}$
and $a \in \CA$. Of course, this action is determined by the bimodule
structure on $\O{1}$ and the definition (\ref{codi}), however,
it remains to be said whether some extra conditions should be assumed.
The only limitation is that the introduced additional restrictions
should reduce to (\ref{codi}) in the case of commutative differential
calculus.

\subsection{Bimodule linear connection}

The proposition that one should use the bimodule isomorphism
$\s$ to define such property, has been put forward by Dubois-Violette,
Madore and others [9-11]:
\beq
D(\om a) \= \s ( \om \tsa da ) + D(\om) a. \l{lcsy}
\eeq
Throughout this paper we shall call connections that use (in any form)
the bimodule property of $\O{1}$ {\em bimodule connections}.
Indeed, this reduces to the standard expression in the classical
case, where $\om a = a \om$ and, since $\s (\om \tsa \rho) = \rho \tsa
\om$ we have (\ref{lcsy}) equivalent to (\ref{codi}). We shall see, however
that this condition is very restrictive in noncommutative case, in fact,
as shown recently \cite{MAD1}, in many cases of noncommutative Kaluza-Klein
theories
thet only existing bimodule linear connections have no mixed terms.
We shall discuss it later while applying the theory to our example of
$\R^n \times \Z_2$.

\subsection{Torsion and curvature}

Let us observe that (\ref{codi}) is, as mentioned earlier, easily extendible
to $\Omega \tsa \O{1}$ according to the rule:
\beq
D( u \tsa \rho) \= du \tsa \rho + (-1)^{\deg u} u \bt D(\rho),
\eeq

where $u \in \Omega$ and $\rho \in \O{1}$. We can calculate then the
curvature $D^2$ and show that it is left-linear:

\beq
D^2 (u \tsa \rho) \= u \bt D( \rho ).
\eeq

Similar extension is not possible for the right-multiplication
property of the covariant derivative, and, what is more important
one cannot assure that the curvature $D^2$ is right-linear.

The torsion could be defined as the following map $T: \Omega \tsa \O{1} \to
\Omega$:
\beq
T \= \pi \circ D - d \circ \pi, \l{tors}
\eeq

where $\pi$ is standard projection. From the construction it is clear
that $T$ is a left-module morphism (in case of symmetric
connections it is a bimodule homomorphism).

Finally, let us make some general observations on linear connections in
noncommutative geometry, which later would be useful.

\bo{3}
If $D$ and $D'$ are two linear connections, then $D-D'$ is a left-linear
morphism $\Omega \tsa \O{1} \to \Omega \tsa \O{1}$ of grade 1, moreover, if
they
are bimodule connections (i.e. obeying (\ref{lcsy}) ) then $D-D'$ is
a bimodule morphism.
\eo

If $\O{1}$ is a free bimodule and $\om_1,\ldots,\om_n$ form its base, then
a connection $D$ such that $D(\om_i)=0$ is called {\em trivial} in this
base. Then, as a result of observation 3 we may observe that in that case
every connection is a sum of this trivial connection and a left-module
(or bimodule in the case of bimodule connections) morphism of grade 1.

To end this section we shall observe that  having a $\star$-structure
on our external algebra we cannot easily relate somehow $D(u)$ with
$D(u^\star)$.
 However, let us observe that in the classical differential geometry
it is not true that $D( \om^\star ) = D (\omega)^\star$

\subsection{$\O{1}$ as a bimodule over $\Omega$}

As we have shown in the previous paragraph, the use of bimodule properties
of linear connection is rather complicated. In what follows we should attempt
to propose a solution, which would make both the notation and results simpler.
The price we have to pay is the introduction of additional structure on our
differential algebra, as we shall assume that there exist a bimodule structure
over $\Omega$ (treated as an algebra) on $\O{1}$. We shall call this bimodule
$\CM$, assuming that the following conditions hold: \\
1. $\CM$ is generated by elements of the form $u \tsa \om$ where $u \in \Omega$
and $\om \in \O{1}$. Of course, $\O{1} \subset \CM$ \\
2. The left- and right-multiplications by the elements of $\Omega$
coincide with  $\tsa$ if the element of the module is in $\O{1}$ and $\bt$
otherwise.\\
3. $\pi: \CM \to \Omega$ defined on the generators $\pi (u \tsa \om) = u \bt
\om$
is bimodule morphism \\
3. There exists a $\star$-operation on $\CM$.

We shall demonstrate that such structure exists in the standard differential
geometry
as well as in few examples of noncommutative geometry. First, let us notice
that having defined this structure we could immediately write both rules for
$D$,
now seen as a map $D: \CM \to \CM$ of degree 1:

\bea
D( u \bt m) & = & du \bt m + (-1)^{\deg u} u \bt D(m) \l{cobi1} \\
D( m \bt u) & = & D(m) \bt u + (-1)^{\deg m} m \bt du \l{cobi2}
\eea
for $m \in \CM$ and $u \in \Omega$.

Now, $D^2$ is automatically a bimodule morphism!.

\subsubsection{Examples}

First, we shall demonstrate that this structure exists in the standard
commutative differential calculus. Define the right action of
$\Omega$ on the generators of $\CM$ as follows:
\beq
\om \tsa u \= (-1)^{\deg u} u \tsa \om,
\eeq
then this gives a proper bimodule structure on $\CM$ and $\pi$ is
a bimodule morphism. We can see that in this case (\ref{cobi1}) is
equivalent to (\ref{cobi2}), as one would expect.

Now let us turn to noncommutative geometry. For universal calculus one
can always introduce the bimodule $\CM$ as $\wedge$ is just $\tsa$ and
$\CM$ could be identified with the tensor algebra of differential
forms itself. For the simplest possible case of two-point geometry we
have:
\bea
D(a \chi ) & = & \chi (\pt a) \tsa \chi + a D(\chi), \\
D(\chi a)  & = & D(\chi) a - \chi \tsa \chi (\pt a),
\eea
and we could verify that they agree with each other provided that
$D(\chi) = 2 \chi \tsa \chi$, so that $D$ coincides with $d$. This
result is what we could have expected, observe that since $\pi$ is
just identity map, every torsion-free connection on universal calculus
must coincide with $d$.

Next we shall discuss the product of continuous and discrete geometries
with the following construction of $\CM$. The bimodule structure on $\CM$
is, for products of the forms $dx^i$, just as in the case of continuous
geometry, as discussed above. Similarly, for products of $\chi$ alone, we
take it as in the example of universal calculus right above.
What we have to add is the rule of right multiplication between
$dx^i$ and $\chi$:
\bea
dx^i \tsa \chi & \sim & - \chi \tsa  dx^i \\
\chi \tsa dx^i & \sim & - dx^i \tsa  \chi
\eea

We could verify now what (\ref{cobi1}-\ref{cobi2}) imply on the
covariant derivative. First let us compare $D( a dx^i)$ with
$D(dx^i a)$:

\bea
D( a dx^i ) = dx^j \tsa dx^i (\pt_j a) +
\chi \tsa dx^i (\pt a) + a D(dx^i), \\
D( dx^i a ) = D(dx^i) a + dx^j \tsa dx^i (\pt_j a) +
\chi \tsa dx^i (\pt a),
\eea

where in the second relation we have used the rules of right
multiplication on $\CM$.  By comparing the right-hand side
of these relations we immediately get that:
\beq
D(dx^i) \= \GH{i}{j}{k} dx^j \tsa dx^k + \alpha \chi \tsa \chi. \l{slc-1}
\eeq

The other pair of relations is:
\bea
D( a \chi ) = dx^i \tsa \chi \CR(\pt_i a) -
\chi \tsa \chi (\pt a) + a D(\chi), \\
D( \chi \CR(a) ) = D(\chi) \CR(a) +  dx^i \tsa \chi \CR(\pt_i a)
+ \chi \tsa \chi (\pt a),
\eea

and by comparing the right-hand sides we get;

\beq
D(\chi) \= 2 \chi \tsa \chi. \l{slc-2}
\eeq

Suppose now that we demand that this connection has a vanishing
torsion (\ref{tors}). We have already observed that (\ref{slc-2}),
which is equal to the connection on $\Z_2$ alone is torsion-free.
For (\ref{slc-1}) the vanishing of torsion is equivalent to $\alpha=0$
and $\GH{i}{j}{k} = \GH{i}{k}{j}$, so in the end we obtain that the
linear connection on $\R^n \times \Z_2$ splits into separate components,
each operating on one element of the product. Therefore, the
curvature has also such property, additionally, as in our case $D$ on
the discrete space is flat $D^2=0$, we have the resulting total
curvature operator to have only the standard contribution coming from
the continuous element of the product. This would suggest that already
on this level, without even introducing the concept of metric connection,
we are certain that for bimodule linear connections there would be no
modifications to gravity, coming from effects of noncommutative geometry
on $\R^n \times \Z_2$.

We shall see, that if we drop the requirement of bimodule property
(in either form) we can proceed with the construction, which shall lead
to some interesting and unexpected features.

\section{Metric linear connections}

In this sections we shall discuss the generalisation of the idea
of metric connections. The form of the definition depends on our
assumptions concerning the bimodule properties of $D$ - as our main
task is to apply the theory to the considered example and we have
already shown that for bimodule connections give no new features in
the theory, we shall concentrate on connection, which
only satisfy (\ref{codi}) alone.

We say that $D$ is {\em metric} if the following holds for
all one-forms $u,v$:

\beq
d g(u, v^\star) \= g( D(u), v^\star ) - g(u, D(v)^\star), \l{meco}
\eeq

where we use a shorthand notation: $g(u_1 \tsa u_2, v) = u_1 g(u_2, v)$.
This definition is well-defined for any $D$ and it gives precise
prescription for metric connection in the commutative limit.

The most general form of torsion-free $D$ is:
\bea
D(dx^\m) & = &\GH{\m}{\n}{\r} dx^{\n} \ts dx^{\r} +
\GT{\m}{\n} \left( dx^\n \ts \chi + \chi \ts dx^\n \right),  \\
D(\chi) & = & 2 \chi \ts \chi + \B2{\m}{\n}  dx^{\m} \ts dx^{\n} +
W_\n \left( dx^\n \ts \chi + \chi \ts dx^\n \right).
\eea

where $\GH{\m}{\n}{\r} = \GH{\m}{\r}{\n}$ and $\B2{\m}{\n} = \B2{\n}{\m}$.
Using the metric (\ref{met1}-\ref{met4}) and the definition (\ref{meco})
we end up with the following set of relations:

\bea
\pt_\r g^{\m\n} & = & \GH{\m}{\r}{\k} g^{\k\n} + \GH{\n}{\r}{\k} g^{\m\k},\\
\pt g^{\m\n} & = & \CR(\GT{\m}{\k}) g^{\k\n} - \CR(g^{\m\k})\GT{\n}{\k},
\label{G-g} \\
\GT{\m}{\n} g & = & g^{\m\k} \B2{\n}{\k} \label{G-b}, \\
0 & = & g^{\m\n} W_\n, \label{im4} \\
\pt_\m g & = &  2 W_\m g, \\
\pt g & = & 0
\eea

and, as we assume that $g^{\m\n}$ is non-degenerate, we immediately get that
$W_\m=0$ and $g=const$. This simplifies the curvature $R=D^2$ and e have:
\bea
R(dx^\mu) & = &  dx^\a \bt dx^\b  \;\;
\left( \pt_\a \GH{\m}{\b}{\g} - \pt_\b \GH{\m}{\a}{\g}
+ \GH{\m}{\a}{\k} \GH{\k}{\b}{\g} - \GH{\m}{\b}{\k} \GH{\k}{\a}{\g}
- \GT{\m}{\a} \B2{\b}{\g} + \GT{\m}{\b} \B2{\a}{\g} \right) \ts dx^\g  \nn \\
 &  & + \;\;  dx^\a \bt \chi  \;\;
 \left( - \pt \GH{\m}{\a}{\g} - \CR( \GH{\m}{\a}{\k} \GT{\k}{\g} )
+ \pt_\a \CR(\GT{\m}{\g}) + \CR(\GT{\m}{\k}) \GH{\k}{\a}{\g} \right)
  \ts dx^\g  \nn \\
&  & + \;\; \chi \bt \chi  \;\;
\left(  2 \GT{\m}{\g}  - \pt \GT{\m}{\g} - \GT{\m}{\k} \CR(\GT{\k}{\g})
\right) \ts dx^\g \nn \\
 &  & + \;\;  dx^\a \bt dx^\b  \;\;
\left( \pt_\a \GT{\m}{\b} - \pt_\b \GT{\m}{\a}
 + \GH{\m}{\b}{\k} \GT{\k}{\a} - \GH{\m}{\a}{\k} \GT{\k}{\b} \right)
\ts \chi  \nn \\
  &  & + \;\;  dx^\a \bt \chi  \;\;
\left( -\pt \GT{\m}{\a} - 2 \CR(\GT{\m}{\a}) + \CR(\GT{\m}{\k})
\GT{\k}{\a} \right) \ts \chi  \nn \\
 &  & \nn \\
R(\chi) & =  &  dx^\a \bt dx^\b  \;\;
\left( \pt_\a \B2{\b}{\g} - \pt_\b \B2{\a}{\g}
+ \GH{\k}{\a}{\g} \B2{\b}{\k} - \GH{\k}{\b}{\g} \B2{\a}{\k} \right)
\ts dx^\g  \nn \\
  &  & + \;\;  dx^\a \bt \chi  \;\;\left( 2 \B2{\a}{\g} - \pt \B2{\a}{\g}
- \CR( \B2{\a}{\k} \GT{\k}{\g} ) \right) \ts dx^\g  \nn \\
  &  & + \;\;  dx^\a \bt dx^\b  \;\;
\left( \GT{\k}{\a} \B2{\b}{\k} - \GT{\k}{\b} \B2{\a}{\k} \right)
\ts \chi  \nn \\
\eea

Now, if we use (\ref{G-b}), we may eliminate $\B2{\m}{\n}$ from
the expressions for $R$. Furthermore, it will be convenient to
use $\t{\m}{\n}$:  $\GT{\m}{\n} = \delta^\m_\n + \t{\m}{\n}$. First,
we may rewrite (\ref{G-g}):
\beq
\CR(\t{\m}{\k}) g^{\k\n} \= \CR(g^{\m\k}) \t{\n\k},
\eeq
or using the inverse $g_{\m\n}$:
\beq
\CR( \t{\k}{\m} g_{\n\k} ) \= \t{\k}{\n} g_{\k\m}. \l{rel4}
\eeq

The curvature tensor, rewritten using only using $\GH{\m}{\n}{\r}$
and $\t{\m}{\n}$ variables (only in the first line we still use
$\GT{\m}{\n}$) is:

\bea
R(dx^\mu) & = & dx^\a \bt dx^\b \;\;
\left( \pt_\a \GH{\m}{\b}{\g} - \pt_\b \GH{\m}{\a}{\g}
+ \GH{\m}{\a}{\k} \GH{\k}{\b}{\g} - \GH{\m}{\b}{\k} \GH{\k}{\a}{\g}
- g g_{\n\g} \left( \GT{\m}{\a}\GT{\n}{\b} - \GT{\m}{\b}\GT{\n}{\a}
\right) \right) \ts dx^\g  \nn \\
 &  & + \;\;  dx^\a \bt \chi  \;\;
 \left( \CR(\t{\m}{\k}) \GH{\k}{\a}{\g} - \CR( \GH{\m}{\a}{\k} \t{\k}{\g} )
+ \pt_\a \CR(\t{\m}{\g} \right)
  \ts dx^\g  \nn \\
&  & + \;\; \chi \bt \chi  \;\;
\left( \delta^\m_\g - \t{\m}{\k} \CR(\t{\k}{\g}) \right) \ts dx^\g \nn \\
 &  & + \;\;  dx^\a \bt dx^\b  \;\;
\left( \pt_\a \t{\m}{\b} - \pt_\b \t{\m}{\a}
 + \GH{\m}{\b}{\k} \t{\k}{\a} - \GH{\m}{\a}{\k} \t{\k}{\b} \right)
\ts \chi  \nn \\
  &  & + \;\;  dx^\a \bt \chi  \;\;
\left( - \delta^\m_\a +  \CR(\t{\m}{\k}) \t{\k}{\a} \right) \ts \chi  \nn \\
 &  & \nn \\
R(\chi) & =  &  dx^\a \bt dx^\b  \;\;
 ( g g_{\m\g} ) \left(
 \pt_\a \t{\m}{\b} - \pt_\b \t{\m}{\a}
 + \GH{\m}{\b}{\k} \t{\k}{\a} - \GH{\m}{\a}{\k} \t{\k}{\b} \right)
\ts dx^\g  \nn \\
  &  & + \;\;  dx^\a \bt \chi \;\;
( g g_{\m\g} )
 \left(  \delta^\m_\a - \t{\m}{\k} \CR(\t{\k}{\a}) \right) \ts dx^\g  \nn \\
  &  & + \;\;  dx^\a \bt dx^\b  \;\;
( g g_{\k\r} )
\left( \t{\k}{\a} \t{\r}{\b} - \t{\k}{\b} \t{\r}{\a} \right)
\ts \chi  \nn
\eea

and we see that some expressions repeat itself in the structure of
the curvature tensor. The Ricci tensor $R_c$ is the trace of the
curvature tensor:

\bea
R_c & = & dx^\b \ts dx^\g \;\;
\left( \pt_\m \GH{\m}{\b}{\g} - \pt_\b \GH{\m}{\m}{\g}
+ \GH{\m}{\m}{\k} \GH{\k}{\b}{\g} - \GH{\m}{\b}{\k} \GH{\k}{\m}{\g}
- g g_{\n\g} \left( \GT{\m}{\m}\GT{\n}{\b} - \GT{\m}{\b}\GT{\n}{\m}
\right) \right) \nn \\
& & - \;\; dx^\b \ts dx^\g \;\; \frac{1}{2}
( g g_{\m\g} )
\left(  \delta^\m_\b - \t{\m}{\k} \CR(\t{\k}{\b}) \right)  \nn \\
& & + \;\; \chi \ts dx^\g\;\; \frac{1}{2}
 \left( \CR(\t{\m}{\k}) \GH{\k}{\m}{\g} - \CR( \GH{\m}{\m}{\k} \t{\k}{\g} )
+ \pt_\m \CR(\t{\m}{\g} \right) \nn \\
& & + \;\; dx^\b \ts \chi \;\;
\CR\left( \pt_\m\t{\m}{\b} - \pt_\b \t{\m}{\m}
 + \GH{\m}{\b}{\k} \t{\k}{\m} - \GH{\m}{\m}{\k} \t{\k}{\b} \right) \nn \\
& & + \;\;  \chi \ts \chi\;\; \frac{1}{2}
\CR\left( - \delta^\m_\m +  \CR(\t{\m}{\k}) \t{\k}{\m} \right)
\eea

and finally the curvature scalar, which is simply the value of the metric
on the Ricci tensor:

\beq
R \=
g^{\b\g} \left( \pt_\m \GH{\m}{\b}{\g} - \pt_\b \GH{\m}{\m}{\g}
+ \GH{\m}{\m}{\k} \GH{\k}{\b}{\g} - \GH{\m}{\b}{\k} \GH{\k}{\m}{\g}
\right) -  g \left( \GT{\m}{\m}\GT{\b}{\b} - \GT{\m}{\b}\GT{\b}{\m}
\right), \l{scalar}
\eeq

Such result is an interesting one - we shall get the action, which
is a sum of two standard Hilbert-Einstein actions for gravity
(one for $g{\mu\nu}(x,+)$ and the other one for $g^{\mu\nu}(x,-)$)
as well as additional terms, which depend only on both metric tensors
(no derivatives!) and the field $\t{\k}{\b}$ (satisfying (\ref{rel4})).
This would suggest that such term plays the role of a constraint, which
enforces relations between $g^{\mu\nu}(x,+)$ and  $g^{\mu\nu}(x,-)$. In
the simplest possible case, when the are equal to each other, it would
reduce itself to the cosmological constant term. This would recover
the results obtained by \cite{KAL,KAS} using different approach based on the
Dirac operator and Wodzicki residue. Further and  more detailed discussion
on the properties of the obtained model of gravity on $\R^n \times \Z_2$
and example solutions shall be presented elsewhere.

\section{Conclusions}

In this paper we have presented few schemes, which have been considered
as a generalisation of linear connections (and related objects) in
noncommutative geometry. Our main aim was to apply these methods to
a simple example of noncommutative Kaluza-Klein type model, being the
product of continuous ($\R^n$) and discrete ($\Z_2$ geometries. Our choice
has been motivated by the interpretation of the electroweak part of the
Standard Model, in which such geometry plays an important role
providing  the explanation of the origin of Higgs field.

We have found that most concepts ale easily translated from
standard differential geometry to the noncommutative case and
give reasonable results in our example. However, some others,
especially the postulate of symmetry imposed on the metric and
bimodule properties of linear connections, can cause rather
significant problems. In particular, in our example each of these
requirements has profound consequences. In the first case, it eliminates
all discrete degrees of freedom in the field theory, whereas in the
second case, it gives no new features of gravity in this setup. Though
the latter may be considered as an acceptable result, we cannot agree
with the former - as, we already know how the theory should look like
\cite{NCG1,JA2}.

Therefore we definitely cannot accept the generalisation of symmetric
metric as discussed here (we still require that is hermitian),
being aware that the other result
might also suggest the second postulate (bilinear linear connections)
goes too far. In our considerations we have also proposed another
version of this postulate, which makes it more natural. One of its
main advantages is that $R$ becomes a bimodule morphism.

On the other hand we have provided a derivation of gravity-type theory
for our example of product geometry, based on the assumption that only
left-linearity is important for linear connections, obtaining quite a
feasible result. \\
\ \\

Of course, it still remains open, whether the accepted methods are
proper for noncommutative geometry, as they are based on what we have
learned from standard differential geometry. The main problem is that
few features, which coincide in the commutative case, are different
if we turn on noncommutativity. One has to choose, which property
is appropriate in such situation and different choices may give
completely different results. It is also not clear why the standard
methods must be followed in noncommutative case, for instance,
we might ask why we have to set the torsion to zero.

We have demonstrated in this paper some good points and problems
of two methods as applied to a simply and - realistic - model. The
results that we have found are important for determination of some
fundamental concepts of noncommutative geometry, however, they have
to be verified using other methods, so that they could be accepted
or properly  generalised for noncommutative geometry, which remains
a big task for future research.
\ \\
\ \\
{\bf Acknowledgements:} The author would like to thank J.Madore
for helpful discussions.

\def\v#1{{ \bf  #1} }

\end{document}